\documentclass{elsart}

\usepackage{graphicx}

\usepackage{amssymb}
\usepackage{amsmath}

\begin{document}
\newcommand{\norm}[1]{\left\Vert#1\right\Vert}
\newcommand{\abs}[1]{\left\vert#1\right\vert}
\newcommand{\set}[1]{\left\{#1\right\}}
\newcommand{\Real}{\mathbb R}
\newcommand{\rd}{{\mathbb R^d}}
\newcommand{\ip}[2]{{#1\cdot #2}}
\newcommand{\eps}{\varepsilon}
\newcommand{\To}{\longrightarrow}
\newcommand{\BX}{\mathbf{B}(X)}
\newcommand{\A}{\mathcal{A}}
\newcommand{\CC}{\mathbb C}

\begin{frontmatter}


\title{Coupled continuous time random walks in finance}

\author{Mark M. Meerschaert\thanksref{ta3}}
\address{Department of Mathematics \& Statistics, University of Otago, Dunedin 9001 New Zealand} \ead{mcubed@maths.otago.ac.nz} \corauth[cor1]{Corresponding
Author}

\author{Enrico Scalas\thanksref{ta1}}
\address{Dipartimento di Scienze e Tecnologie Avanzate, Universit\`a del Piemonte Orientale, Alessandria, Italy} \ead{scalas@unipmn.it} 

\thanks[ta3]{M. M. Meerschaert was partially supported by NSF grants
DMS-0139927 and DMS-0417869 and Marsden grant UoO 123.}

\thanks[ta1]{E. Scalas was partially supported by the Italian M.I.U.R. F.I.S.R. Project
``Ultra-high frequency dynamics of financial markets'' and by the EU COST
P10 Action, ``Physics of Risk''.}


\begin{abstract}
Continuous time random walks (CTRWs) are used in physics to model anomalous diffusion, by incorporating a random waiting time between particle jumps.  In finance, the particle jumps are log-returns and the waiting times measure delay between transactions.  These two random variables (log-return and waiting time) are typically not independent.  For these coupled CTRW models, we can now compute the limiting stochastic process (just like Brownian motion is the limit of a simple random walk), even in the case of heavy tailed (power-law) price jumps and/or waiting times.  The probability density functions for this limit process solve fractional partial differential equations.  In some cases, these equations can be explicitly solved to yield descriptions of long-term price changes, based on a high-resolution model of individual trades that includes the statistical dependence between waiting times and the subsequent log-returns. In the heavy tailed case, this involves operator stable space-time random vectors that generalize the familiar stable models.  In this paper, we will review the fundamental theory and present two applications with tick-by-tick stock and futures data.
\end{abstract}

\end{frontmatter}

Continuous time random walk (CTRW) models impose a random waiting time between particle jumps.  They are used in statistical physics to model anomalous diffusion, where a cloud of particles spreads at a rate different than the classical Brownian motion, and may exhibit skewness or heavy power-law tails.  In the coupled model, the waiting time and the subsequent jump are dependent random variables.  See Metzler and Klafter \cite{MetzlerK,MetzlerKlafter} for a recent survey.  Continuous time random walks are closely connected with fractional calculus.  In the classical random walk models, the scaling limit is a Brownian motion, and the limiting particle densities solve the diffusion equation. The connection between random walks, Brownian motion, and the diffusion equation is due to Bachelier \cite{Bachelier} and Einstein \cite{Einstein}.  Sokolov and Klafter \cite{SokolovKlafter} discuss modern extensions to include heavy tailed jumps, random waiting times, and fractional diffusion equations.  

In Econophysics, the CTRW model has been used to describe the movement of log-prices \cite{GMSR,MRGS,RSM,SKDR,SGM}.  An empirical study of tick-by-tick trading data for General Electric stock during October 1999 (Figure 1, left) in Raberto, et al. \cite{RSM} uses a Chi-square test to show that the waiting times and the subsequent log returns are not independent.  These data show that long waiting times are followed small (in absolute value) returns, while large returns follow very short waiting times.  This dependence seems intuitive for stock prices, since trading accelerates during a time of high volatility \cite{betram04}. LIFFE  bond futures from September 1997 (Figure 1, right) show a different behavior, where long waiting times go with large returns.  See \cite{RSM} for a detailed description of the data.  In both cases, it seems clear that the two variables are dependent.  In the remainder of this paper, we will describe how coupled continuous time random walks can be used to create a high-resolution model of stock prices in the presence of such dependence between waiting times and log-returns.  We will also show how this fine scale model transitions to an anomalous diffusion limit at long time scales, and we will describe fractional governing equations that can be solved to obtain the probability densities of the limiting process, useful to characterize the natural variability in price in the long term.     

\begin{figure}\label{GEplot}
\begin{center}
\includegraphics[scale=0.35]{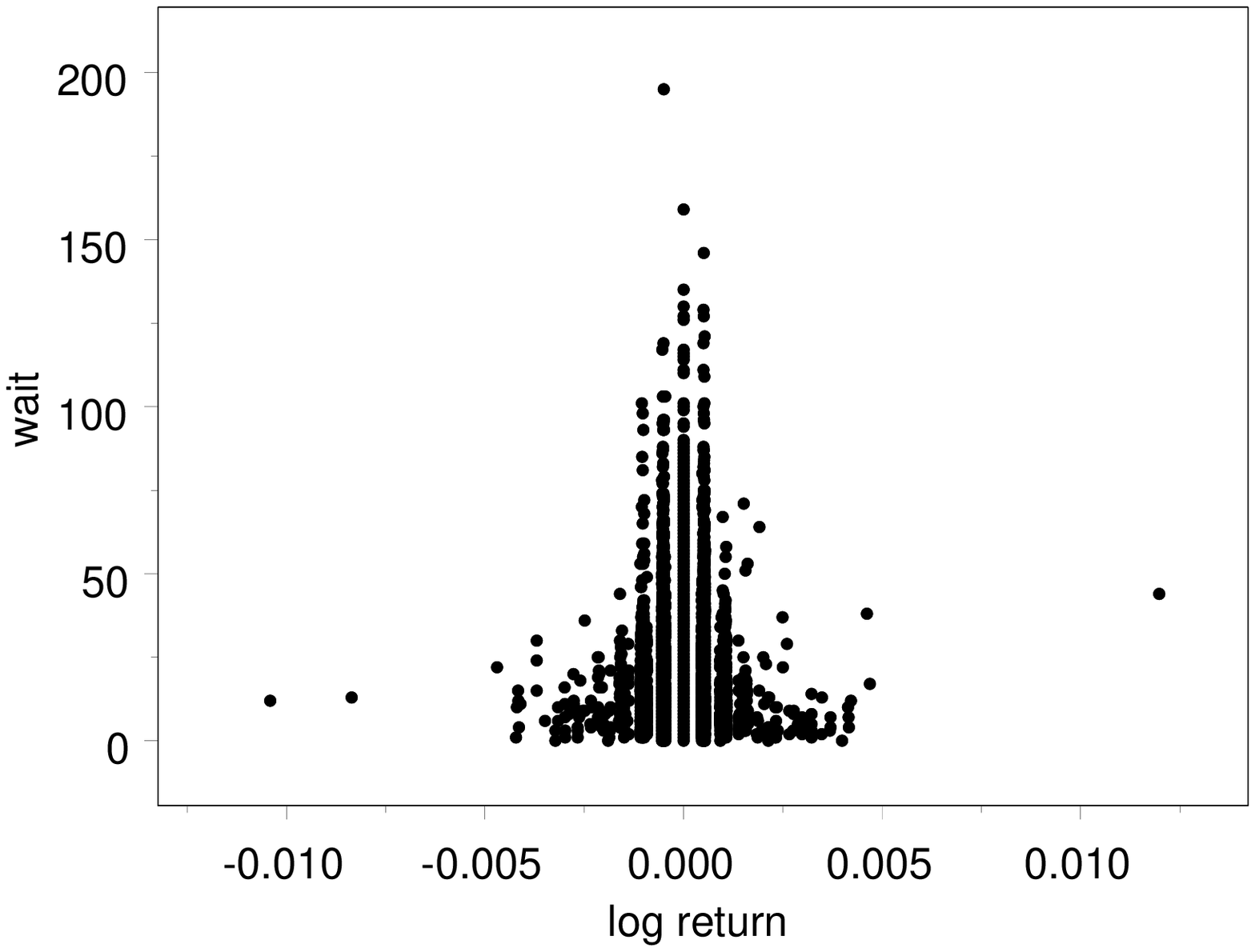}
\includegraphics[scale=0.35]{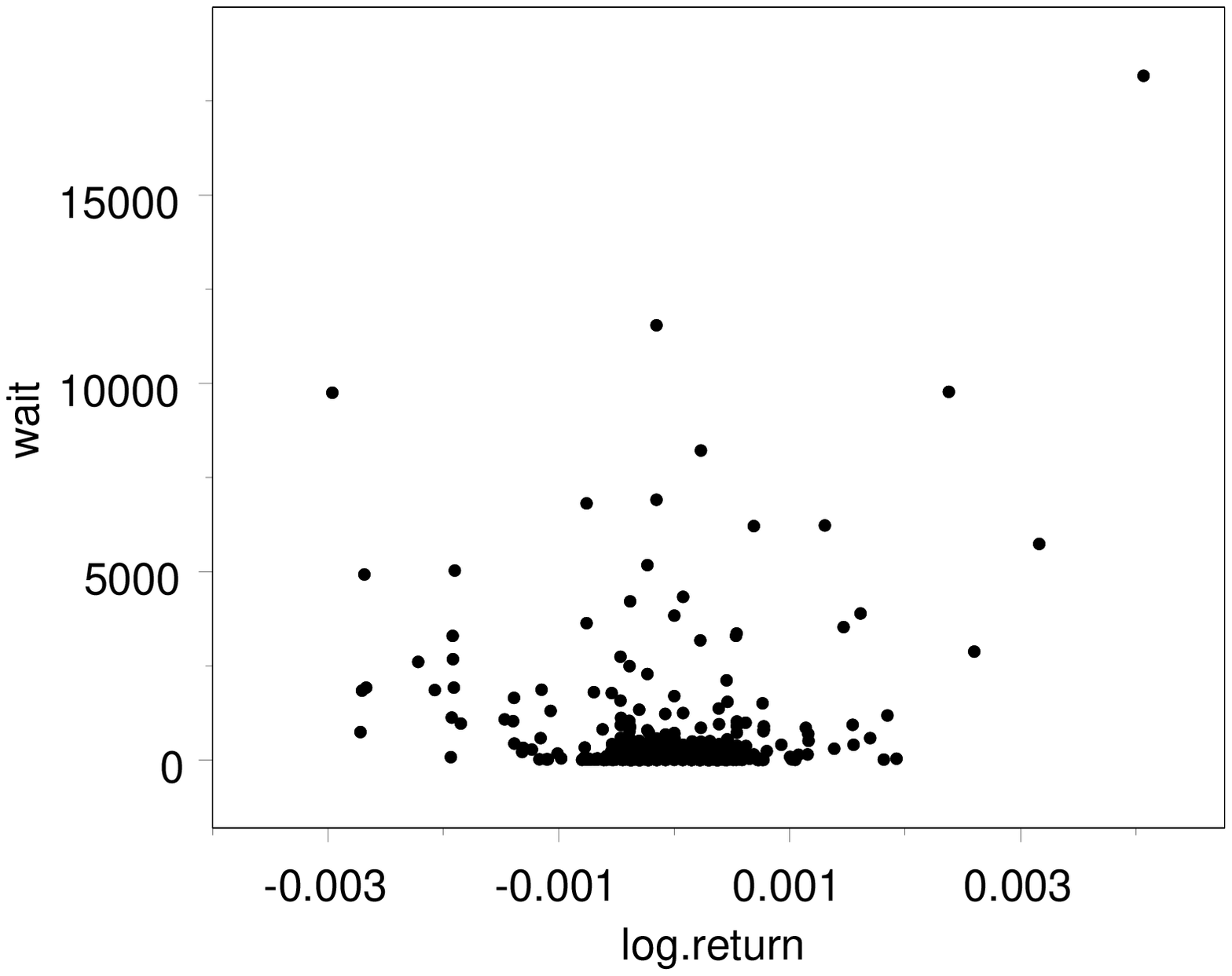}
\caption{Waiting times in seconds and log returns for General Electric stock (left) and LIFFE bond futures (right) show significant statistical dependence.}
\label{fig1}
\end{center}
\end{figure} 

Let $P(t)$ be the price of a financial issue at time $t$.  Let $J_1, J_2, J_3, \ldots$ denote the waiting times between trades, assumed to be nonnegative, IID random variables.  Also let $Y_1, Y_2, Y_3, \ldots$ denote the log-returns, assumed to be IID.  We specifically allow that $J_i$ and $Y_i$ are coupled, i.e., dependent random variables for each $n$.  Now the sum $T_n=J_1+\cdots+J_n$ represents the time of the $n$th trade.  The log-returns are related to the price by $Y_n=\log[P(T_n)/P(T_{n-1})]$ and the log-price after $n$ trades is $S_n=\log[P(T_n)]=Y_1+\cdots+Y_n$.  The number of trades by time $t>0$ is $N_t=\max\{n:T_n\leq t\}$, and the log-price at time $t$ is $\log P(t)=S_{N_t}=Y_1+\cdots+Y_{N_t}$.  

The asymptotic theory of continuous time random walk (CTRW) models describes the behavior of the long-time limit.  For more details see \cite{coupleCTRW,Zsolution,couplePR}.  The log-price $\log P(t)=S_{N_t}$ is mathematically a random walk subordinated to a renewal process.  If the log-returns $Y_i$ have finite variance then the random walk $S_n$ is asymptotically normal.  In particular, as the time scale $c\to\infty$ we have the stochastic process convergence $c^{-1/2}S_{[ct]}\Rightarrow A(t)$, a Brownian motion whose densities $p(x,t)$ solve the diffusion equation $\partial p/\partial t=D \partial^2 p/\partial x^2$ for some constant $D>0$ called the diffusivity.  If the waiting times $J_i$ between trades have a finite mean $\lambda^{-1}$ then the renewal theorem \cite{Feller} implies that $N_t\sim \lambda t$ as $t\to\infty$, so that $S_{N_t}\approx S_{\lambda t}$, and hence the CTRW scaling limit is still a Brownian motion whose densities solve the diffusion equation, with a diffusivity proportional to the trading rate $\lambda$.  If the symmetric mean zero log-returns have power-law probability tails $P(|Y_i|>r)\approx r^{-\alpha}$ for some $0<\alpha<2$ then the random walk $S_n$ is asymptotically $\alpha$-stable, and $c^{-1/\alpha}S_{[ct]}\Rightarrow A(t)$ where the long-time limit process $A(t)$ is an $\alpha$-stable L\'evy motion whose densities $p(x,t)$ solve a (Riesz-Feller) fractional diffusion equation $\partial p/\partial t=D \partial^\alpha p/\partial |x|^\alpha$.  If the waiting times have power-law probability tails $P(J_i>t)\approx t^{-\beta}$ for some $0<\beta<1$ then the random walk of trading times $T_n$ is also asymptotically stable, with $c^{-1/\beta}T_{[ct]}\Rightarrow D(t)$ a $\beta$-stable L\'evy motion.  Since the number of trades $N_t$ is inverse to the trading times (i.e., $N_t\geq n$ if and only if $T_n\leq t$), it follows that the renewal process is asymptotically inverse stable $c^{-\beta}N_{ct}\Rightarrow E(t)$ where $E(t)$ is the first passage time when $D(\tau)>t$.  Then the log-price $\log P(t)=S_{N_t}$ has long-time asymptotics described by $c^{-\beta/\alpha}\log P(ct)\Rightarrow A(E_t)$ a subordinated process.  If the waiting times $J_i$ and the log-returns $Y_i$ are uncoupled (independent) then the CTRW scaling limit process densities solve $\partial^\beta p/\partial t^\beta=D \partial^\alpha p/\partial |x|^\alpha+\delta(x) t^{-\beta}/\Gamma(1-\beta)$ using the Riemann-Liouville fractional derivative in time. This space-time fractional diffusion equation was first introduced by Zaslavsky \cite{Zaslavsky,Zaslavsky2} to model Hamiltonian chaos.  Explicit formulas for $p(x,t)$ can be obtained via the inverse L\'evy transform of Barkai \cite{Barkai,Zsolution} or the formula in \cite{mainardi01}. 

If the waiting times $J_i$ and the log-returns $Y_i$ are coupled (dependent) then the same process convergence holds, but now $E(t)$ and $A(t)$ are not independent.  Dependent CTRW models were first studied by Shlesinger et al. \cite{SKW} in order to place a physically realistic upper bound on particle velocities $Y_i/J_i$.  They set $Y_i=J_i^{\beta/\alpha} Z_i$ where $Z_i$ is independent of $J_i$.  In their example, they assume that $Z_i$ are independent, identically distributed normal random variables, but the choice of of $Z_i$ is essentially free \cite{coupleCTRW}.  Furthermore, any coupled model at all for which $(c^{-1/\alpha}S_{[ct]},c^{-1/\beta}T_{[ct]})\Rightarrow (A(t),D(t))$ will have one of two kinds of limits:  Either the dependence disappears in the limit (because the waiting times $J_n$ and the log-returns $Y_n$ are asymptotically independent), or else the limit process is one of those obtainable from the Shlesinger model \cite{coupleCTRW}.  In the former case, the long-time limit process densities are governed by the space-time fractional diffusion equation of Zaslavsky \cite{scagormaimee,scalas1,scalas2}.  In the remaining case, the long-time limit process densities solve a coupled fractional diffusion equation $\left({\partial}/{\partial t}-{\partial^\alpha}/{\partial |x|^\alpha}\right)^\beta p(x,t)=\delta(x){t^{-\beta}}/{\Gamma(1-\beta)}$ with $\alpha=2$ in the case where $Z_i$ is normal \cite{couplePR}.  In that case, the exact solution of this equation is 
\begin{equation}\label{CTRWdensityeq}
p(x,t)=\int_0^t \frac{1}{\sqrt{4\pi u}}\exp\left(-\frac{x^2}{4u}\right)\frac{u^{\beta-1}}{\Gamma(\beta)}\frac{(t-u)^{-\beta}}{\Gamma(1-\beta)}\ du
\end{equation}
which describes the probability distributions of log-price in the long-time limit.  The resulting density plots are similar to a normal but with additional peaking at the center, see Figure 2 (right).

\begin{figure}\label{CoupleDensity}
\hskip-10pt\includegraphics[scale=0.275]{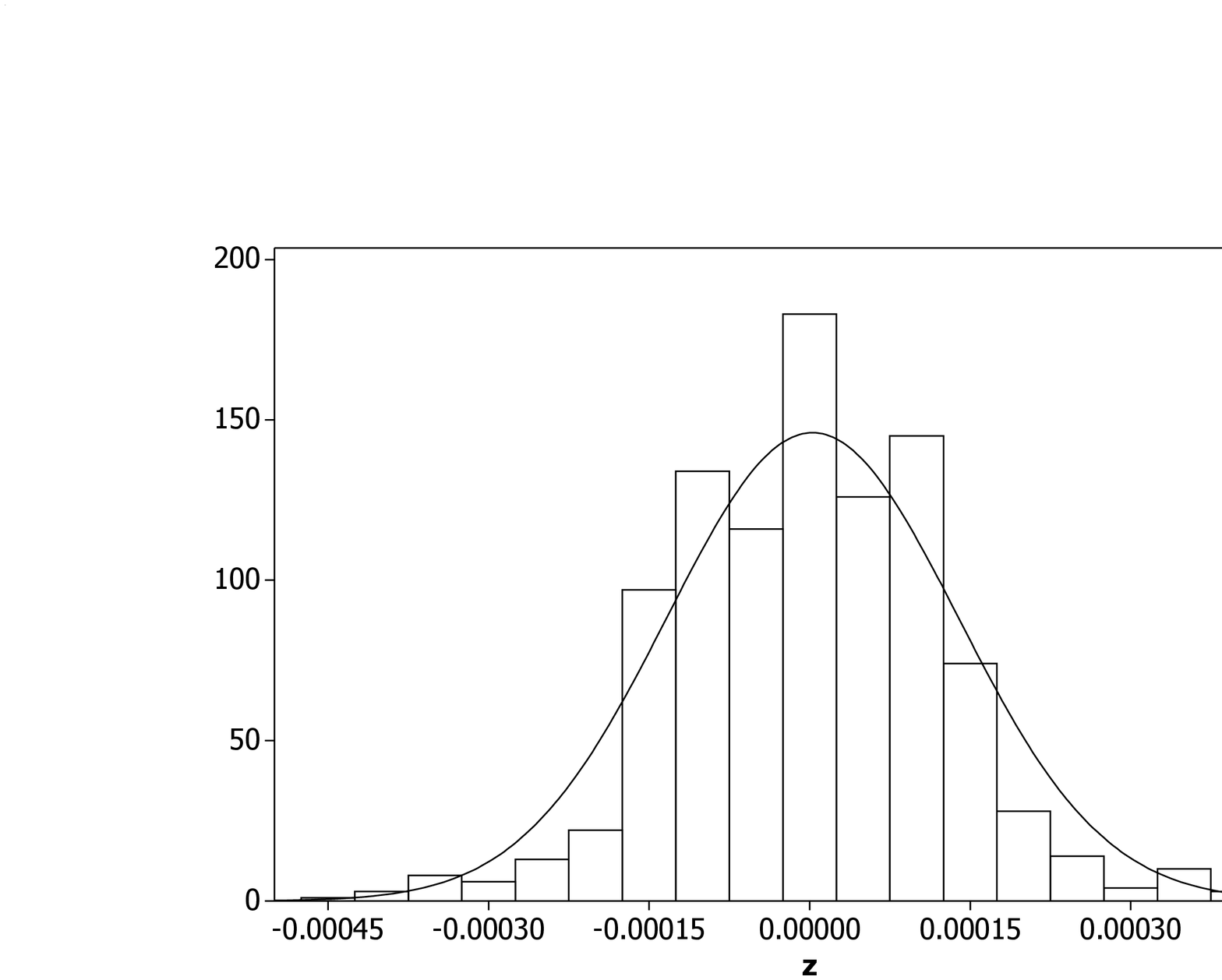}
\hskip-10pt\includegraphics[scale=0.275]{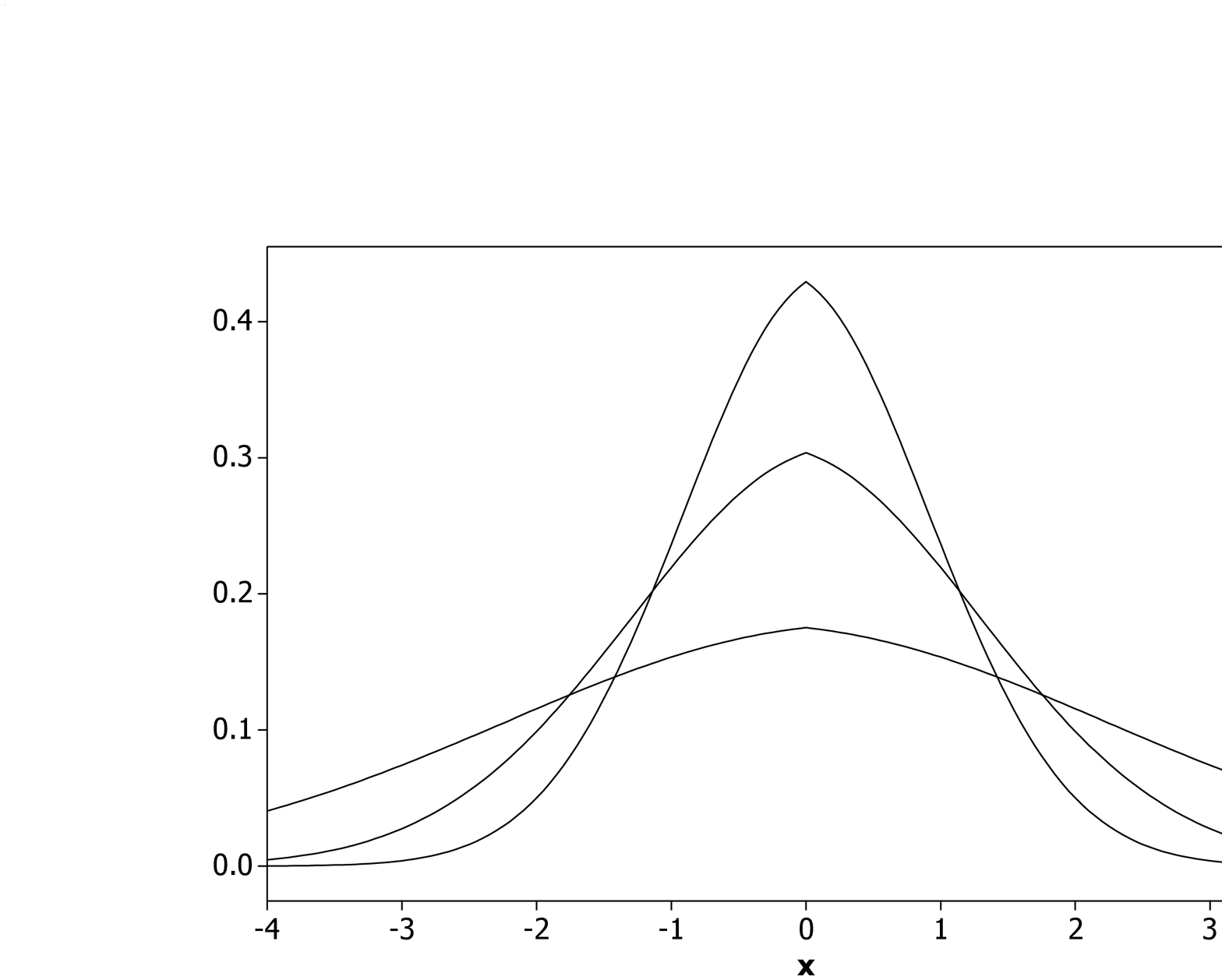}
\vskip-80pt
\caption{Coupled CTRW model for LIFFE futures using normal coupling variable (left) produces limit densities (right) from equation \eqref{CTRWdensityeq} for $t=0.5, 1.0, 3.0$.  }
\end{figure}

As noted above, even if the waiting times and log-returns are dependent, it is possible that the dependence disappears in the long-time limit.  The relevant asymptotics depend on the space-time random vectors $(T_n,S_n)$, which are asymptotically operator stable \cite{RVbook}.  In fact we have the vector process convergence $(c^{-1/\beta}T_{[ct]},c^{-1/\alpha}S_{[ct]})\Rightarrow (D(t),A(t))$ and it is possible for the component processes $A(t)$ and $D(t)$ of this operator stable L\'evy motion to be independent.  The asymptotics of heavy tailed random vectors (or random variables) depend on the largest observations \cite{portfolio} and hence the general situation can be read off Figure 1.  When components are independent, the largest observations cluster on the coordinate axes.  This is because the rare events that cause large waiting times or large absolute log-returns are unlikely to occur simultaneously for both, in the case where these two random variables are independent.  Hence we expect a large value of one to occur along with a moderate or small value of the other, which puts these data points far out on one or the other coordinate axis.  If the components are only asymptotically independent, the same behavior will be seen on the scatterplot for the largest outlying values, even though the two coordinates are statistically dependent.  This is just what we see in Figure 1 (left), and hence we conclude that for the GE stock, the coupled CTRW model has exactly the same long-time behavior as the uncoupled model analyzed previously \cite{GMSR}.  

The coupling $Y_i=J_i^{\beta/\alpha} Z_i$ in the Shlesinger model implies that the longest waiting times are followed by large log-returns.  For the data set shown in Figure 1 (right), it is at least plausible that the Shlesinger model holds.  To check this, we computed $Z_i=J_i^{-\beta/\alpha} Y_i$ for the largest 1000 jumps, following the method of \cite{Scheffler99}.  We estimated $\alpha=1.97$ and $\beta=0.95$ using Hill's estimator.  The ``size" of the random vector $(J_i,Y_i)$ is computed in terms of the Jurek distance $r$ defined by $(Y_i,J_i)=(r^{1/\alpha}\theta_1, r^{1/\beta}\theta_2)$ where $\theta_1^2+\theta_2^2=1$ \cite{portfolio}.  The resulting data set $Z_i$ can be adequately fit by a normal distribution (see Figure 2, left).  Hence the Shlesinger model provides a realistic representation for the coupled CTRW in this case.  To address a slight lack of fit at the extreme tails, we also experimented with a centered stable with index 1.8, skewness 0.2, and scale 0.08 (not shown), where the parameters were found via the maximum likelihood procedure of Nolan \cite{Nolan}.  For the stable model, the long-time limit densities can be obtained by replacing the normal density in equation \eqref{CTRWdensityeq} with the corresponding stable density.  

In summary, we have shown that the coupled-CTRW theory can be applied to
finamcial data. We have presented two different data sets, GE Stocks
traded at NYSE in 1999, and LIFFE bond futures from 1997. In both cases
there is statistical dependence between log-returns and waiting time, but
the asymptotic behaviour is different leading to different theoretical
descriptions.

\end{document}